\documentclass{webofc}
\pdfoutput=1
\usepackage[varg]{txfonts}
\usepackage{listings}
\usepackage{tikz}
\usepackage{color}
\usepackage{url}
\usepackage{array}
\usepackage{footnote}
\usepackage{titlesec}
\usepackage[bottom]{footmisc}
\usepackage[]{hyperref}

\DeclareFixedFont{\ttb}{T1}{txtt}{bx}{n}{12}
\DeclareFixedFont{\ttm}{T1}{txtt}{m}{n}{12}

\definecolor{keycolor}{RGB}{58,129,195}
\definecolor{emphcolor}{RGB}{78,49,99}
\definecolor{strcolor}{RGB}{45,149,116}
\definecolor{bkgcolor}{gray}{0.98}

\newcommand\pythonstyle{
  \lstset{
    language=Python,
    otherkeywords={self},
    keywordstyle=\color{keycolor},
    emph={All,Any},
    emphstyle=\color{emphcolor},
    stringstyle=\color{strcolor},
    showstringspaces=false
}}

\lstnewenvironment{python}[1][]
{
\pythonstyle
\lstset{#1}
}
{}

\newcolumntype{L}[1]{>{\raggedright\let\newline\\\arraybackslash\hspace{0pt}}p{#1}}
\newcolumntype{C}[1]{>{\centering\let\newline\\\arraybackslash\hspace{0pt}}p{#1}}
\newcolumntype{R}[1]{>{\raggedleft\let\newline\\\arraybackslash\hspace{0pt}}p{#1}}

\makesavenoteenv{tabular}

\hypersetup{
  pdftitle={AlphaTwirl},
  pdfauthor={Tai Sakuma},
  bookmarksnumbered=true,
  bookmarksopen=true,       
  bookmarksopenlevel=1,
  colorlinks=true,
  allcolors=blue
}

\begin{document}

\graphicspath{{images/}}
\DeclareGraphicsExtensions{.pdf,.png}

\title{AlphaTwirl}
\subtitle{A Python library for summarizing event data \\
  into multivariate categorical data}

\author{\firstname{Tai}
  \lastname{Sakuma}\inst{1}\fnsep\thanks{\email{tai.sakuma@bristol.ac.uk}}}

\institute{University of Bristol}

\abstract{AlphaTwirl is a Python library that summarizes large event
  data into multivariate categorical data, which can be regarded as
  generalizations of histograms. The output can be imported as data
  frames in R and pandas. With their rich set of data wrangling
  tools, users can develop flexible and configurable analysis code.
  The multivariate categorical data loaded as data frames are readily
  visualized by graphic tools available in R and Python. AlphaTwirl
  can process event data concurrently with multiple cores or batch
  systems. Users can extend and customize nearly any functionality of
  AlphaTwirl with reusable code. AlphaTwirl is released under the BSD
  license.}

\maketitle
\section{Introduction}
\label{sec:intro}

\begin{figure}[!b]
\centering
\includegraphics[width=0.75\textwidth]{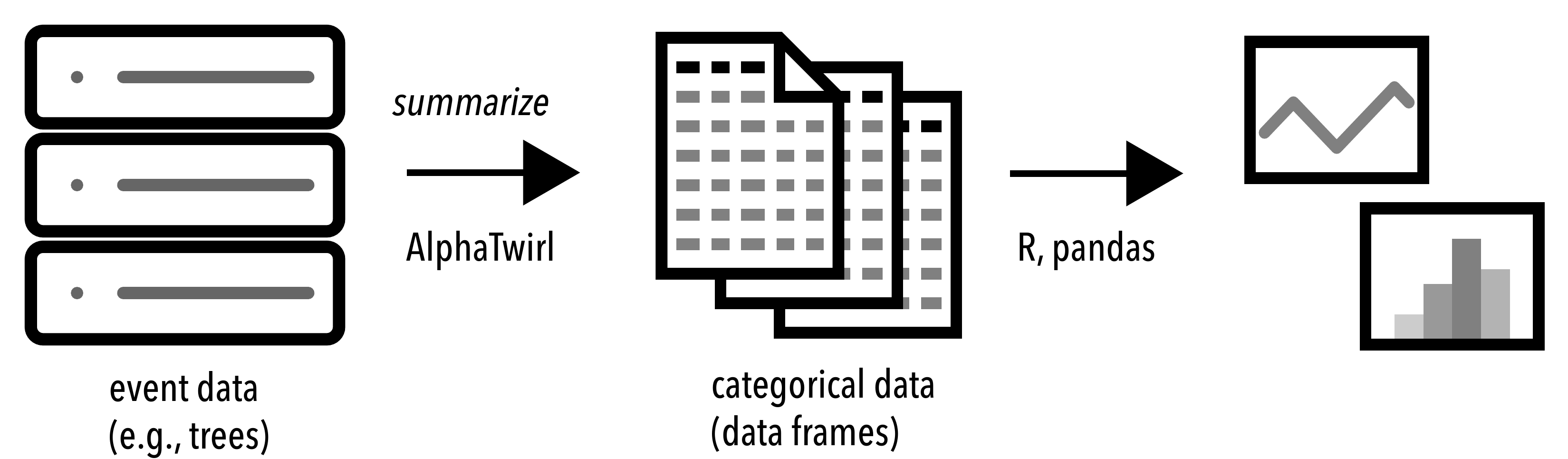}
\label{fig:workflow}
\caption{Large event data are summarized by AlphaTwirl into
  categorical data, which are imported as data frames in R and
  pandas.}
\end{figure}

\textit{AlphaTwirl} is a Python library that summarizes large
\textit{event data} into a set of \textit{multivariate categorical
  data}, which can be loaded as \textit{data frames} in R~\cite{r} and
pandas~\cite{pandas-scipy-2010}, as depicted in
Fig.~\ref{fig:workflow}. AlphaTwirl is used in the CMS
experiment~\cite{Chatrchyan:2008aa} to analyze event data in ROOT
trees~\cite{Brun:1997pa}, including Delphes
trees~\cite{deFavereau:2013fsa}, Heppy trees~\cite{heppy}, and CMS
MiniAOD~\cite{Petrucciani:2015gjw} and NanoAOD~\cite{nanoaod}.
AlphaTwirl enabled the development of new dimensionless variables for
\mbox{supersymmetry} searches~\cite{Sakuma:2018xrq}. AlphaTwirl is
available at Ref.~\cite{alphatwirlgithub} under the BSD license.

This paper starts by distinguishing event data and categorical data,
followed by the discussion of how data frames with categorical data
can be regarded as generalizations of histograms and their advantages.
The paper, then, describes how AlphaTwirl summarizes event data as
well as how it selects events and adds variables on the fly. The paper
also mentions features of the implementation, such as dependency
injection, framework independent modules, and concurrency.

\section{Event data and categorical data}

Event data and categorical data can be distinguished as follows.

\begin{description}
\item[Event data.] Event data are any data with one entry per event.
  Events can be any kind from coin tosses, to bank transactions, to
  proton-proton collisions at the LHC. Data in ROOT trees are
  typically event data. Event data can be very large because they have
  as many entries as the number of the events. Event data are often
  stored on dedicated storage systems.
\item[Categorical data.] Categorical data are any data with one entry
  per category, for example, histograms. Categorical data are
  summaries of event data. They are usually small because they only
  have as many entries as the number of the categories---small enough
  to be loaded into memory on a laptop computer.
\end{description}

\section{Data frames as generalizations of histograms}
\label{sec:df}

\textit{Data frames} are a common data structure in R and pandas. Data
frames, in their usual form, have a two-dimensional tabular structure:
different \textit{columns} contain different variables, different
\textit{rows} different entries. Data frames can express both event
data\footnote{Event data with multiple variable-length lists, e.g., a
  list of jets and a list of muons, can be expressed by data frames in
  several ways; for example, (a) by a data frame with as many rows per
  event as the length of the longest lists in the event, or (b) by
  multiple data frames---one data frame per list, as common in
  relational databases.} and categorical data.

Histograms are a widely used data structure in high energy physics. In
fact, ROOT has dedicated classes for histograms, such as TH1D and
TH2D. Histograms are a special case of categorical data. Histograms
are lists of the sums of weights for each category, where categories
are specified, for example, by one or a combination of ordered labels,
unordered labels, discrete variables, and intervals of discrete or
continuous variables.

Data frames can express histograms, as in the following example:
\vspace{0.2\baselineskip}
\begin{lstlisting}[
    escapechar=\#,
    xleftmargin=.15\textwidth, xrightmargin=.15\textwidth,
    framexleftmargin=12pt, framexrightmargin=12pt
  ]
process    ht njet   met         n      nvar
    QCD   400    2   200  4.44e+05  2.83e+07
    QCD   400    2   220  2.96e+05  1.68e+07
    QCD   400    2   240  1.90e+05  9.83e+06
                       #\vdots#
 TTJets  1200    6  1180  1.43e-01  1.49e-02
 TTJets  1200    6  1200  7.09e-01  9.07e-02
\end{lstlisting}
\vspace{0.2\baselineskip}
The example data frame is a histogram of yields of proton-proton
collision events at the LHC evaluated in Monte Carlo simulation in
intervals of three kinematic variables (\lstinline{ht},
\lstinline{njet}, \lstinline{met}) for each generated process such as
QCD. The data frame has six columns. The first four columns
(\lstinline{process}, \lstinline{ht}, \lstinline{njet},
\lstinline{met}) specify categories---four dimensional categories. The
\lstinline{process} is unordered labels; the other three dimensions,
\lstinline{ht}, \lstinline{njet}, and \lstinline{met}, are intervals
of numerical variables. In this example, the intervals are represented
by their lower edges.\footnote{The locations of upper edges can be
  stored in data frames in many different ways. One method is to use
  lower edges as bin labels as in the example and only include all
  non-empty bins and the next bins of all non-empty bins in a data
  frame. The next bins of all non-empty bins are included so that the
  upper edges of all non-empty bins can be found. Notice that upper
  edges of empty bins have no information about data.} The data frame
has as many rows as the categories, that is, the unique combinations
of all possible values in the first four columns. The other two
columns (\lstinline{n}, \lstinline{nvar}) show, respectively, the sums
of the weights and the sums of the squares of the weights; namely,
\lstinline{n} and \lstinline{nvar} in each row are summaries of the
events in the category specified by the first four columns.

As can be seen from the above example, data frames can express
arbitrary dimensions of histograms with categories specified by
combinations of different types of variables such as strings,
integers, and floats. In addition, data frames can have an arbitrary
number of summary columns. For example, data frames can include
columns for \textit{means}, \textit{minima}, and \textit{maxima} of
numerical variables.\footnote{A summary can also be a variable-length
  array, in which case, a data frame will have as many rows per
  category as the length of the array for the category.} In short,
histograms are summaries of event data and can be generalized to
multivariate categorical data, which can be expressed as data frames.

\section{Advantages of data frames---data wrangling, visualization}

Once event data are summarized into multivariate categorical data,
they can be imported as data frames in R and pandas, both of which
offer a rich set of data wrangling tools: sort by values, concatenate,
merge by keys, reshape between long and wide formats, and so on. In
particular, the \textit{split-apply-combine} strategy~\cite{plyr} has
many applications. These tools make it easy to write flexible and
configurable analysis code, which, for example, can help users quickly
try many different analysis methods to exploit the event data.

Furthermore, multivariate categorical data in data frames in R are
readily visualized by \textit{lattice}~\cite{lattice},
\textit{ggplot2}~\cite{ggplot2}, and other R visualization packages;
data frames in pandas can be similarly visualized by
\textit{seaborn}~\cite{seaborn} and other Python packages. These
visualization tools can, for example, let users quickly investigate
how events are distributed in a multidimensional kinematic phase
space.

\section{Why does AlphaTwirl summarize event data?}

Why does AlphaTwirl summarize event data? In other words, why not
simply convert the data type of event data to data frames?

It is usually possible to convert the data type to data frames. For
example, there are tools to convert ROOT trees to data frames of R and
pandas, such as \textit{RootTreeToR}~\cite{roottreetor} and
\textit{root\_pandas}~\cite{rootpandas}. However, event data are often
too large for R and pandas because they both usually load all data
into memory. In fact, a set of event data analyzed for a single
publication in CMS is not nearly small enough to fit in memory on a
typical desktop or laptop computer even in the smallest format of
NanoAOD. Consequently, it is not straightforward to fully benefit from
the data wrangling and visualization tools mentioned in the previous
section on a whole set of event data even if their data types are
somehow converted to data frames.

On the other hand, the next step after the conversion in many analyses
in high energy physics, in particular binned analyses, would be to
make histograms, or, in more general terms, summarize the event data
into categorical data. For these reasons, AlphaTwirl summarizes event
data as it loads them\footnote{Consequently, a summary that requires
  access to data from multiple or all events simultaneously to
  compute, such as \textit{medians}, is not trivial to implement.}
rather than converting the data type.

\section{Summarizing event data by split-apply-combine strategy}

Event data can be summarized into multivariate categorical data as a
data frame by the split-apply-combine strategy~\cite{plyr}:

\vspace{0.5\baselineskip}
\noindent\makebox[\textwidth][c]{%
\begin{minipage}{0.7\textwidth}
\textbf{split:} split event data into groups determined as categories.

\textbf{apply:} apply a function to summarize the data in each group.

\textbf{combine:} combine the results as a data frame.
\end{minipage}}

\vspace{0.5\baselineskip}

\noindent
In fact, histograms can be created by this strategy---split data into
bins, sum the weights in each bin, and combine the results. With
\textit{dplyr}~\cite{dplyr} in R or the function \lstinline{groupby()}
of pandas, this strategy can be easily used for small event data. An
aim of AlphaTwirl is to summarize large event data as they stream by
the split-apply-combine \mbox{strategy}.\footnote{The new ROOT class
  \textit{RDataFrame}~\cite{rdataframe}, which first appeared in ROOT
  v6.14/00 in 2018, can summarize large ROOT trees by a similar
  strategy; that is, it can aggregate data in trees by a user-defined
  accumulation operation.}

\vspace{\baselineskip}
\noindent
\textbf{Code example.} Users can specify how to summarize event data
by a Python dictionary. The following code is a simple example of
creating a two-dimensional histogram:
\begin{python}[
    xleftmargin=.05\textwidth, xrightmargin=.05\textwidth,
  ]
  dict(key_name=('ht', 'jet_pt'), key_index=(None, 0),
       key_binning=(Binning(boundaries=(400, 800, 1200)),
                    RoundLog(0.1, 100)))
\end{python}
This simple example only specifies groups into which to \textit{split}
data. The example does not specify a summarizing function to
\textit{apply} or a method to \textit{combine}, letting the default
determine them; the default summarizing function counts the number of
the entries in each group and the results are by default combined into
a data frame, that is, making a data frame with an unweighted
histogram. The above example dictionary has three entries. Their
values are tuples with the same length:
\begin{center}
\begin{tabular}{rL{0.69\textwidth}}
{\small\ttfamily key\_name}: & Names of categorical variables or
variables whose intervals are categorical variables, {\small\ttfamily
  ht} and {\small\ttfamily jet\_pt} in the example. They can be branch
names if the input data is a ROOT tree. They can also be names of
variables created on the fly by \textit{scribblers}, described in
Section~\ref{sec:scribblers}. \\

{\small\ttfamily key\_index}: & Indices to be used for each variable
in the {\small\ttfamily key\_name} if the variable is an array. In the
example, the first index is {\small\ttfamily None} because the first
variable {\small\ttfamily ht} is not an array.\footnote{Internally,
  scalar variables are implemented as Python lists with single
  elements. The index~{\scriptsize\ttfamily 0} can be, therefore, used
  as well instead of {\scriptsize\ttfamily None}.} The second variable
{\small\ttfamily jet\_pt} is an array. The second index,
{\small\ttfamily 0}, will be its index, i.e., {\small\ttfamily
  jet\_pt[0]}. Indices of arrays can be flexibly specified as
described below. \\

{\small\ttfamily key\_binning}: & Functions to place values into
intervals for each variable in the {\small\ttfamily key\_name}. They
can be any functions that take a value and return a bin name. The
{\small\ttfamily Binning} and {\small\ttfamily RoundLog} are functors
included in AlphaTwirl. The {\small\ttfamily Binning} functor uses the
specified bin boundaries. The {\small\ttfamily RoundLog} functor uses
bins with an equal width in the log scale. The first argument
{\small\ttfamily 0.1} is the width in the log scale and the second
argument {\small\ttfamily 100} is a boundary; therefore, the bin
boundaries are $\cdots, 10^{1.9}, 10^{2} , 10^{2.1}, 10^{2.2},
\cdots$.\footnote{Unless optionally specified, the smallest and
  largest boundaries are determined by the system, for example,
  {\scriptsize\ttfamily sys.float\_info.min} and {\scriptsize\ttfamily
    sys.float\_info.max}.}
\end{tabular}
\end{center}
If summary methods require variables to be summarized, for example,
taking the mean of a variable, the variable names and indices can be
specified with, respectively, \lstinline{val_name} and
\lstinline{val_index} in similar ways to \lstinline{key_name} and
\lstinline{key_index}.

\vspace{\baselineskip}
\noindent
\textbf{Indices of arrays---wildcards, back references.} Indices of
arrays can be flexibly specified with \textit{wildcards} and
\textit{back references}. For example, a four-dimensional categorical
variable of $p_\text{T}$ and $\eta$ of all possible pairs of a jet and
a muon can be specified as follows.
\begin{python}[
    xleftmargin=.05\textwidth, xrightmargin=.05\textwidth,
  ]
  dict(key_name=('jet_pt', 'jet_eta', 'muon_pt', 'muon_eta'),
       key_index=('(*)', '\\1', '(*)', '\\2'), ...)
\end{python}
The syntax is inspired by the regular expression. The wildcard
\lstinline{'*'} indicates all elements of the array. If
\lstinline{key_index} contains multiple wildcards, all possible
combinations of the elements from each array are used. The back
reference \lstinline{'\\n'} indicates the same index as the n-th
wildcard within parentheses.
 
\section{Event selections and graph theory}
\label{sec:selection}

Some events may not be needed. Conditions of event selections can
be flexibly specified as nested conditions combined with the logical
conditions, \lstinline{All}, \lstinline{Any}, and \lstinline{Not}. An
example code is as follows.
\begin{python}[
    xleftmargin=.05\textwidth, xrightmargin=.05\textwidth,
  ]
  dict(All=('ev: ev.ht[0] >= 400',
             dict(Any=('ev: ev.njet[0] == 1',
                  dict(All=('ev: ev.njet[0] >= 2',
                            'ev: ev: ev.mht[0] >= 200'))
  ))))
\end{python}
A condition can be specified by a string or dictionary. A string such
as \lstinline{'ev:} \lstinline{ev.ht[0]} \lstinline{>=}
\lstinline{400'} will be parsed and executed as a Python lambda, where
the argument \lstinline{ev} will be the \textit{event object}, which
will be described in Section~\ref{sec:eventloop}. Each dictionary has
one entry with the key \lstinline{'All'}, \lstinline{'Any'}, or
\lstinline{'Not'}. If the key is \lstinline{'All'} or
\lstinline{'Any'}, the value is a tuple of conditions. If the key is
\lstinline{'Not'}, the value is a condition. If the key is
\lstinline{'All'}, events need to satisfy all conditions in the tuple.
If the key is \lstinline{'Any'}, events need to satisfy at least one
condition in the tuple. If the key is \lstinline{'Not'}, events should
not satisfy the condition.

Event selections are implemented as \textit{directed trees} with each
condition as a vertex. For example, the above example code corresponds
to the following graph.
\begin{center}
  \begin{tikzpicture}
  [scale=.8,auto=left,every node/.style={rectangle,fill=gray!20}]
  \node (n1) at (4,4) {\lstinline{All}};
  \node (n2) at (2,3)  {\lstinline{ev.ht[0] >= 400}};
  \node (n3) at (6,3)  {\lstinline{Any}};
  \node (n4) at (4,2)  {\lstinline{ev.njet[0] == 1}};
  \node (n5) at (8,2)  {\lstinline{All}};
  \node (n6) at (6,1)  {\lstinline{ev.njet[0] >= 2}};
  \node (n7) at (10,1)  {\lstinline{ev.mht[0] >= 200}};
  \foreach \from/\to in {n1/n2,n1/n3,n3/n4,n3/n5,n5/n6,n5/n7}
    \path[->] (\from) edge (\to);
  \end{tikzpicture}
\end{center}
Each vertex has one incoming edge except the one at the top. The
logical conditions \lstinline{All} and \lstinline{Any} can have any
number of ordered outgoing edges. The logical condition
\lstinline{Not} (not used in the example) has one outgoing edge. The
conditions specified by strings have no outgoing edges
(\textit{leaves}). Event data can be summarized at leaves.
\textit{Scribblers}, described in the next section, can be placed at
leaves as well.

The graph implementation makes it easier to add functionalities in
vertices and edges. In fact, AlphaTwirl includes two sets of
implementations of the logical conditions \lstinline{All},
\lstinline{Any}, and \lstinline{Not}. While one only evaluates the
conditions themselves, the other counts the number of events
satisfying each condition connected to outgoing edges, which, for
example, can be used to generate selection efficiency
(\textit{cutflow}) tables. Furthermore, users can provide their own
implementations of the logical conditions with desired functionalities
at runtime.

\section{\textit{Scribblers}---adding variables}
\label{sec:scribblers}

If input event data do not contain variables used for selecting or
summarizing events, they can be created on the fly by
\textit{scribblers}. AlphaTwirl does not include a scribbler.
Scribblers are usually provided by users.

A collection of scribblers can be found in Ref.~\cite{scribblers}. It
includes scribblers that apply \textit{NumPy}~\cite{numpy} functions
to arrays, form objects from arrays with the same length, flatten
objects to arrays, select objects by the graph implementation
described in the previous section, correct objects by a given
function, and match objects based on a distance calculated by a given
function. These scribblers can, for example, be used to form a list of
jet objects from arrays such as \lstinline{jet_pt},
\lstinline{jet_eta}, apply jet energy corrections, and flatten back to
multiple arrays.

CMS EDM trees~\cite{cmsedm} contain persistent objects of classes. In
such a case, scribblers need to be developed to unpack persistent
objects into primitive types and their arrays.

\section{Implementation features}
\label{sec:features}

\noindent
\textbf{Dependency injection.} Classes in AlphaTwirl generally operate
on arguments of their methods (\textit{duck typing})---a code example
in the next section. Therefore, the actual implementations of nearly
all functionality are determined at runtime and can be provided by the
user. Examples from the previous sections include binning functors,
logical conditions of event selections, and scribblers.

\vspace{0.5\baselineskip}

\noindent
\textbf{Framework independent modules.} Conversely, each particular
implementation does not normally depend on AlphaTwirl either.
Therefore, they can be reused in a different framework with simple
adapters. For example, the graph implementation of event selections
can be reused in Heppy. Users can extend, customize, and replace
almost any functionality with reusable code at runtime.

%

\vspace{0.5\baselineskip}

\noindent
\textbf{Concurrency.} Large event data can be split into chunks and
processed concurrently with multiple cores or batch systems.
AlphaTwirl includes code to use Python
\textit{multiprocessing}~\cite{multiprocessing} and
HTCondor~\cite{condor-practice}. Users can provide code to use other
systems at runtime; for example, the SGE system can be used with the
code in Ref.~\cite{atsge}. Input data can be split in terms of the
numbers of files and events. While jobs are running in background
processes or a batch system, the main process is running in the
foreground, monitoring the progress of the jobs, and collecting the
results as jobs finish. Failed jobs are automatically resubmitted.

\section{Event readers, event objects, and event loops}
\label{sec:eventloop}

This section describes how operations and features discussed in the
previous four sections work together with a code excerpt.
Schematically, the class \lstinline{EventLoop} is implemented as
follows.
\begin{python}[
    xleftmargin=.15\textwidth, xrightmargin=.15\textwidth,
  ]
  class EventLoop:
      def __init__(self, build_events, reader):
          self.build_events = build_events
          self.reader = reader
     def __call__(self):
          events = self.build_events()
          self.reader.begin(events)
          for ev in events:
              self.reader.event(ev)
          self.reader.end()
          return self.reader
\end{python}
Instances of this class are the event loops that are dispatched to
background processes or to worker nodes of batch
systems\footnote{Because of the dispatch, {\scriptsize\ttfamily
    build\_events} and {\scriptsize\ttfamily reader} need to be
  serializable (\textit{picklable}). The object {\scriptsize\ttfamily
    events} does not need to be serializable. The
  {\scriptsize\ttfamily reader} does not need to stay serializable
  after it {\scriptsize\ttfamily begin}s until it
  {\scriptsize\ttfamily end}s.} and executed concurrently.
\textit{Event loops} loop over events and have \textit{event readers}
read each event. Event readers can be scribblers, objects that select
events, objects that summarize event data, or \textit{composites} of
these in the \textit{composite pattern}~\cite{designpatterns}. Event
readers in AlphaTwirl do not need inherit from any particular base
class, which makes event readers portable as discussed in the previous
section. The object \lstinline{reader} in the code above can be a
single event reader, a composite of event readers, or the object at
the top of the directed tree described in Section~\ref{sec:selection}.
Each \lstinline{EventLoop} instance loops over a different set of
events. Typically, the iterable object \lstinline{events} or its
iterator is connected to input data files and, at each iteration,
loads data for one event into the \textit{event object}
\lstinline{ev}.

The class \lstinline{EventLoop} simply operates on the objects it
initially receives---just as a feature described in the previous
section---and does not depend on the natures of events or the reader.
They can be determined at runtime. AlphaTwirl includes the iterable
events that load event data from ROOT trees with only primitive types
such as integers and floats and arrays of those.\footnote{While PyROOT
  is used in the implementation, branches are accessed by addresses
  rather than attributes of the tree object for the speed.} Iterable
events that load from Delphes trees and CMS EDM trees can be found,
respectively, in Ref~\cite{atdelphes} and Ref~\cite{atcmsedm}. The
package \textit{atuproot}~\cite{atuproot} includes iterable events
that load many events to the event object at each iteration from ROOT
trees by using columnar access of \textit{uproot}~\cite{uproot}, which
can be used together with event readers that can read multiple events
at a time. It is also possible to develop iterable events that load
data from sources other than ROOT files.

\section{Summary}

AlphaTwirl summarizes large event data into multivariate categorical
data by the split-apply-combine strategy. The indices of input
variables can be flexibly specified with wildcards and back
references. Event selections are implemented as directed trees with
each condition as a vertex, at which functionalities can be
implemented. New variables can be added on the fly by scribblers.
Nearly any functionality can be extended or customized with reusable
code. AlphaTwirl can concurrently process event data with multiple
cores or batch systems.

Multivariate categorical data can be regarded as generalizations of
histograms and imported as data frames in R and pandas, which makes it
easier to develop flexible and configurable analysis code with data
wrangling tools in R and pandas. Multivariate categorical data
imported as data frames are readily visualized by graphic tools
available in R and Python, which for example help users visually
inspect event distributions in a multidimensional \mbox{phase space}.

\begin{flushleft}
  \bibliography{proceedings}
\end{flushleft}

\end{document}